\pdfoutput=1
\documentclass[12pt,colorlinks,breaklinks,allcolors=blue!60!black]{iopart}
\usepackage[english]{babel}
\newcommand{\newblock}{}
\expandafter\let\csname equation*\endcsname\relax
\expandafter\let\csname endequation*\endcsname\relax

\usepackage{mathtools}
\usepackage{amsfonts}
\usepackage{graphicx}
\usepackage[utf8]{inputenc}
\usepackage{lmodern}
\usepackage[style=numeric-comp,url=false,backend=biber,%
firstinits=true,sorting=none,clearlang=true,%
maxnames=3,minnames=1,uniquename=false,%
maxbibnames=10
]{biblatex}
\usepackage{xcolor}
\usepackage{hyperref}
\usepackage{csquotes}
\usepackage{microtype}
\usepackage{amsthm}

\usepackage{booktabs}
\usepackage{paralist}
\usepackage{siunitx}
\newcommand{\Text}[1]{\text{\textnormal{#1}}}

\newtheorem{theorem}{Theorem}

\DeclareRobustCommand{\inlinelist}[1]{\begin{inparaenum}[a)] #1 \end{inparaenum}}

\begin{document}

\title{Energy-time entanglement, Elements of Reality, and Local
Realism}


\author[J. Jogenfors]{Jonathan Jogenfors}
\ead{jonathan.fors@liu.se}
\address{Institutionen för
systemteknik, Linköpings Universitet, 581 83 Linköping, SWEDEN}
\author[J.-Å. Larsson]{Jan-Åke Larsson}
\ead{jan-ake.larsson@liu.se}

\address{Institutionen för
systemteknik, Linköpings Universitet, 581 83 Linköping, SWEDEN}

\begin{abstract}
	The Franson interferometer, proposed in 1989 [J.~D. Franson,
	\emph{Phys. Rev. Lett.} \textbf{62}:2205--2208 (1989)], beautifully
	shows the counter-intuitive nature of light. The quantum description
	predicts sinusoidal interference for specific outcomes of the
	experiment, and these predictions can be verified in experiment. In
	the spirit of Einstein, Podolsky, and Rosen it is possible to ask if the
	quantum-mechanical description (of this setup) can be considered
	complete. This question will be answered in detail in this paper,
	by delineating the quite complicated relation between energy-time
	entanglement experiments and Einstein-Podolsky-Rosen (EPR) elements
	of reality. The mentioned sinusoidal interference pattern is the
	same as that giving a violation in the usual Bell experiment. Even
	so, depending on the precise requirements made on the local realist
	model, this can imply
	\inlinelist{
	\item no violation,
	\item smaller violation than
		usual, or
	\item full violation of the appropriate statistical
		bound.
	}
	Alternatives include
	\inlinelist{
	\item using only the measurement outcomes
		as EPR elements of reality,
	\item using the emission time as EPR
		element of reality,
	\item using path realism, or
	\item using a modified setup.
	}
	This paper discusses the nature of these alternatives and
	how to choose between them. The subtleties of this discussion needs
	to be taken into account when designing and setting up experiments
	intended to test local realism. Furthermore, these considerations
	are also important for quantum communication, for example in
	Bell-inequality-based quantum cryptography, especially when aiming
	for device independence.
\end{abstract}

\maketitle

\section{Introduction}
\label{sec:introduction}

In 1989 a new interferometric setup was proposed by J.~D.~Franson~\cite{Franson89}.
The main intent was to test the possibility of local realist
models as a possible description, more complete than quantum mechanics. The
sinusoidal interference obtained from the experiment when restricting to
coincident events is larger than the bound from given by the Bell inequality~\cite{Bell64}. But the selection of coincident events at the two sites
introduces postselection into the data analysis. This need for postselection has
been under discussion for some time~\cite{deCaro94,AKLZ,Franson00,Ryff01,AKLZ01,Franson09}, and this paper is
intended to review the discussion and to provide some insight into the matter at
hand. We will see that, depending on what is required from the tested model
class, the appropriate inequality changes so that the same experimental outcomes
in some cases do violate the Bell inequality as usual, and in some cases do not.
Interestingly, the class of models that uses EPR elements of reality (and
nothing more) falls between these two, and violates the Bell inequality at a
lesser degree than in other setups.

The paper is organized as follows: the rest of the Introduction is devoted to
background, in a level of detail that enables an in-depth discussion in what
follows. Section\nobreakspace \ref {sec:franson} introduces the Franson interferometer and discusses
effects of postselection. In section\nobreakspace \ref {sec:pathrealism} the usual Bell inequality is
re-established by adding path realism to the model class.
Section\nobreakspace \ref {sec:localrealism} concentrates on using only EPR elements of reality,
giving a weaker inequality but nonetheless a violation of local realism, and
section\nobreakspace \ref {sec:modified-setups} contains a few examples of modified experimental
setups and their properties.

A central concept in this analysis is \enquote{EPR elements of reality} as
proposed by Einstein, Podolsky, and Rosen (EPR) in 1935~\cite{EPR}. The concept
is well-known, but a brief repetition is in place. The setting is as follows:
consider a (small) physical system on which we intend to measure position $Q$ or
momentum $P$. The physical measurement devices associated with these
measurements are mutually exclusive, and furthermore, the quantum-mechanical
description for this physical system tells us that the measurements $Q$ and $P$
do not commute. The standard way to interpret this is that the system does not
possess the properties of position or momentum, only probabilities are possible
to obtain from quantum mechanics.

What EPR ask in their paper is whether the quantum-mechanical description can be
considered complete, or if it is possible to argue for another, more complete
description. They use a combined system of two subsystems $A$ and $B$ of the
above type in a combined state, so that measurement of the sum of the positions
gives $Q_A+Q_B=0$ and measurement of the momentum difference gives $P_A-P_B=0$.
These two combinations are measurable at the same time even though the
individual positions and momenta are not, which means that it is possible to
produce a joint state with these properties. Letting the two subsystems
separate, usually very far, they consider individual measurement of position or
momentum. The system is such that the position sum and momentum difference is
preserved under the separation process, which means that if the position of one
subsystem has been measured, the position of the remote subsystem can be
predicted. Therefore, EPR argue, the position of the remote subsystem must exist
as a property of the subsystem. EPR write:

\begin{figure*}[t]
	\centering
	\includegraphics[width=\linewidth]{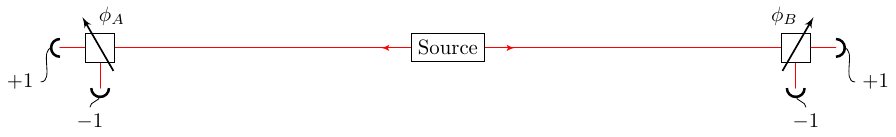}
	\caption{The EPR-Bohm-Bell setup. The two systems are spin-1/2
		systems, and the local measurements are made along a direction
		in space $\phi_A$ or $\phi_B$, respectively. The source is such that
		if the directions $\phi_A=\phi_B$, the outcomes $A+B=0$ with
	probability one.}\label{fig:epr-bohm}
\end{figure*}

\begin{quote}
	If, without in any way disturbing a system, we can predict with
	certainty (i.e., with a probability equal to unity) the value of a
	physical quantity, then there exists an element of physical reality
	corresponding to this physical quantity.
\end{quote}

Likewise, if the momentum of one subsystem has been measured, the
momentum of the remote subsystem can be predicted. In this case, the
momentum of the remote subsystem must exist as a property of the
subsystem. EPR continue to argue that \emph{both} position and
momentum must simultaneously exist as properties of the remote
subsystem, otherwise

\begin{quote}
	\ldots the reality of $P$ and $Q$ depend[s] upon the process of
	measurement carried out on the first system, which does not disturb
	the second [remote] system in any way. No reasonable definition of
	reality could be expected to permit this.
\end{quote}

The above definition of an \enquote{EPR element of reality} is the
philosophical motivation for considering properties of a system as
existing, independent of measurement. The possibility of remote
prediction (\enquote{without in any way disturbing a system}) enables the
notion of EPR element of reality, and we will use this notion below to
motivate existence of properties of the involved physical systems.

A measurement setup more suited to experiment was proposed by
\citeauthor{Bohm1951} in the fifties~\cite{Bohm1951} (see figure\nobreakspace \ref {fig:epr-bohm}),
and uses a system combined of two spin-1/2 subsystems in a total spin 0 state,
so that the spins $A+B=0$ when measured along equal directions $\phi_A=\phi_B$.
The subsystems are allowed to separate and a spin measurement is made on one of
the subsystems. The choice of measurement directions is a continuous choice
instead of the dichotomic choice in the original EPR setup. When a measurement
has been made on one subsystem, the result can be used to predict the result of
a measurement on the remote subsystem along the same direction. And because the
reality of the spin measurement result in the remote system cannot depend on the
local choice, the spin along \emph{any} direction is an EPR element of reality.

This was used in the celebrated Bell paper~\cite{Bell64} where a
statistical test was devised in the form of an inequality that must
be fulfilled by any mathematical model that is realist and
local. Here, realism is motivated by the spin being an EPR element of
reality, and locality by the finite speed of light, or
more specifically, because local measurement is made \enquote{without in any
way disturbing} the remote system (the formulation below is adapted
from~\cite{Jalar98a}).

\begin{theorem}\label{thm:localrealism}
	A local realist model has the following two
	properties:
	\begin{description}
		\item [Realism.] Measurement outcomes can be described by two families
			of random variables. $A$ is the outcome for site 1 with local setting $\phi_A$ and
			$B$ the outcome for site 2 with local setting $\phi_B$:
			\begin{equation*}
				A(\phi_A,\phi_B,\lambda)\Text{ and }B(\phi_A,\phi_B,\lambda)
			\end{equation*}
			where the absolute values of the outcomes are bounded by 1.
			The dependence on the hidden variable $\lambda$ is usually suppressed in the notation

		\item [Locality.] Outcomes do not depend on the remote settings
			\begin{equation*}
				A(\phi_A,\phi_B,\lambda)=A(\phi_A,\lambda),\quad
				B(\phi_A,\phi_B,\lambda)=B(\phi_B,\lambda).
			\end{equation*}
	\end{description}
	By writing $A_i=A(\phi_{A,i})$ and $B_j=B(\phi_{B,j})$ the outcomes from a local realist model obey
	\begin{equation}\label{eq:bell}
		\Big|E\big(A_1B_1\big)
		+E\big(A_1B_2\big)\Big|
		+\Big|E\big(A_2B_1\big)
		-E\big(A_2B_2\big)\Big| \le2.
	\end{equation}
\end{theorem}
Inequality~(\ref {eq:bell}) is violated by the predictions of quantum
mechanics, for instance measurement on a state with total spin zero
gives the correlation
\begin{equation}
	\label{eq:spin-correlation}
	\big\langle A(\phi_A)B(\phi_B)\big\rangle=-\cos(\phi_A-\phi_B)
\end{equation}
with $\phi_A-\phi_B$ being the angle between the two directions $\phi_A$ and
$\phi_B$. Choosing the four directions $\pi/4$ apart in a plane in the
order $b_1$, $a_1$, $b_2$ and $a_2$ one obtains
\begin{equation}
	\label{eq:qm-value}
	\Big|\big\langle A_1B_1\big\rangle +\big\langle
	A_1B_2\big\rangle \Big|
	+\Big|\big\langle
	A_2B_1\big\rangle -\big\langle
	A_2B_2\big\rangle \Big| =2\sqrt2,
\end{equation}
and this violates inequality~(\ref {eq:bell}). Therefore, the
quantum-mechanical predictions cannot be obtained from a local realist
model.

There is one problem that is present in most experiments done to test
this inequality: inefficient detectors. This problem was first noticed
by \citeauthor{Pearle} in 1970~\cite{Pearle} but has been treated in many
subsequent papers. The problem is that only pairs with coincident detections are
obtained, which is a subset of all pairs emitted by the source. The correlation obtained from experiment
is conditioned on coincident detection of two particles, one at each
side. Taking this into account, the theorem needs to be modified as
follows (adapted from~\cite{Jalar98a}).

\begin{theorem}\label{thm:inefficiency}
	A local realist model with inefficiency has the
	following three properties:
	\begin{description}
		\item[Realism.] Outcomes are given by random variables on subsets of
			detection (in a probability space $(\Lambda,\mathcal F,P)$),
			\begin{align*}
				A(\phi_A,\phi_B,\lambda)\Text{ on } \Lambda_{A,\phi_A,\phi_B}
				\quad\Text{ and }\quad
				B(\phi_A,\phi_B,\lambda)\Text{ on } \Lambda_{B,\phi_A,\phi_B}.
			\end{align*}

		\item[Locality.] Outcomes and detections do not depend on the remote
			settings,
			\begin{align*}
				A(\phi_A,\phi_B)=A(\phi_A)& \quad \Text{ on } \quad
				\Lambda_{A,\phi_A,\phi_B}=\Lambda_{A,\phi_A}\\
				B(\phi_A,\phi_B)=B(\phi_B)& \quad \Text{ on } \quad
				\Lambda_{B,\phi_A,\phi_B}=\Lambda_{B,\phi_B}.
			\end{align*}

		\item[Efficiency.] There is a lower bound to the efficiencies,
			\begin{equation*}
				\eta=\underset{\parbox{14mm}{\centering\scriptsize{\upshape
				settings\\local sites}}}{\min} P\Big(\Text{coincidence}
				\Big|\Text{local detection}\Big).
			\end{equation*}
	\end{description}
	The outcomes from a local realist model with
	inefficiency obey
	\begin{equation}\label{eq:bell-inefficiency}
		\begin{split}
			\Big|&E\big(A_1B_1\big|\Text{coinc. for }A_1\Text{ and }B_1\big)
			+E\big(A_1B_2\big|\Text{coinc. for }A_1\Text{ and }B_2\big)\Big|\\
			&+\Big|E\big(A_2B_1\big|\Text{coinc. for }A_2\Text{ and }B_1\big)
			-E\big(A_2B_2\big|\Text{coinc. for }A_2\Text{ and }B_2\big)\Big|\le\frac4\eta-2.
		\end{split}
	\end{equation}
\end{theorem}

The effect is that the inequality is weakened by inefficient detectors, and is
no longer violated by quantum mechanics at $\eta=2\sqrt2-2\approx82.83\%$. We
will see effects and extensions of this below.

\section{The Franson interferometer}\label{sec:franson}

\begin{figure*}[t]
	\centering
	\includegraphics[width=\linewidth]{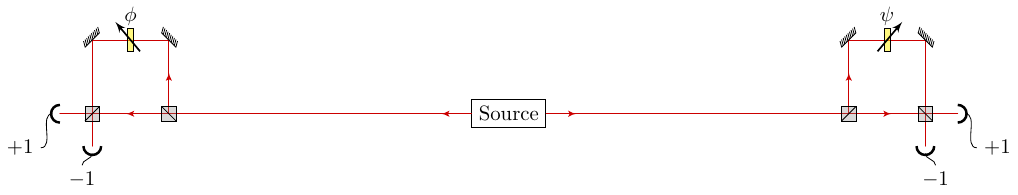}
	\caption{The Franson setup. The source sends out time-correlated
		photons at unknown moments in time. These travel through
		analysis stations consisting of unbalanced (but equal) Mach-Zehnder interferometers with variable
		phase modulators $\phi_A$ and $\phi_B$. If the detections are coincident
	and $\phi_A+\phi_B=0$, then $A=B$ with probability one.}\label{fig:franson}
\end{figure*}

In 1989 a new experimental setup was proposed by \citeauthor{Franson89}~\cite{Franson89} (see figure\nobreakspace \ref {fig:franson}). The setup uses a source that emits
time-correlated photons at unknown moments in time, and two analysis stations
consisting of unbalanced Mach-Zehnder interferometers. The analysis stations
should have a path difference that is large enough to prohibit first-order
interference. Therefore, the probability is equal for a photon to emerge from
each port of the final beamsplitter. The path differences of the analysis
stations should be as equal as possible; the path-difference difference
(repetition intended) should be so small that the events of both photons
\enquote{taking the long path} and both photons \enquote{taking the short path}
are indistinguishable. Quotation marks are used here to remind the reader that
photons are not particles, but quantum objects and as such, do not take a
specific path. Given this indistinguability, since the emission time is unknown
(and is a quantum variable in second quantization), there can be interference
between two possibilities at the second beamsplitter.

There will be no interference if one photon \enquote{takes the long path} and
the other \enquote{takes the short}, because then the emission time can be
calculated easily as the early detection time minus the short path length
divided by $c$. When this happens the emission time will become known. When both
photons \enquote{take the same path}, the emission time remains unknown, and
this is what enables the interference. The interference is not visible as a
change in output intensity, as in first-order interference, but instead in
correlation of the outputs. Given coincident detection, if the total phase
modulation
$\phi_A+\phi_B=0$, then a photon emerging in the $+1$ channel on the left is
always accompanied by a photon emerging in the $+1$ channel on the right, and
the same for the $-1$ channels. Therefore, when a measurement has been made at
one analysis station the result can be used to predict what port the photon will
emerge from at the remote analysis station whenever $\phi_A+\phi_B=0$. Since
the local phase modulation can be chosen freely, the output port along \emph{any}
direction is an EPR element of reality, when coincidence occurs. As a function
of the total phase modulation, we have

\begin{equation}
	\label{eq:cosine-correlation}
	\big\langle A(\phi_A)B(\phi_B)\big|
	\Text{coinc. for }A(\phi_A)\Text{ and }B(\phi_B)\big\rangle=\cos(\phi_A+\phi_B).
\end{equation}
Not the similarity to equation\nobreakspace \textup {(\ref {eq:spin-correlation})}. It is now simple to obtain
\begin{equation}\label{eq:franson-qm}
	\begin{split}
		\Big|\big\langle
		A_1&B_1\big|\Text{coinc. for }A_1\Text{ and }B_1\big\rangle +\big\langle
		A_1B_2\big|\Text{coinc. for }A_1\Text{ and }B_2\big\rangle \Big|\\
		&+\Big|\big\langle
		A_2B_1\big|\Text{coinc. for }A_2\Text{ and }B_1\big\rangle -\big\langle
		A_2B_2\big|\Text{coinc. for }A_2\Text{ and }B_2\big\rangle \Big| =2\sqrt2,
	\end{split}
\end{equation}
which exceeds the bound in inequality~(\ref {eq:bell}). Does this mean that the Franson
interferometer violates local realism?

The answer is no. The problem is that there exists a local
realist model that gives the exact same predictions as quantum
mechanics~\cite{AKLZ} (see figure\nobreakspace \ref {fig:lhv}). Since the model fulfulls the
requirements of Theorem\nobreakspace \ref {thm:localrealism}, it is strange that it seems to
violate inequality~(\ref {eq:bell}). But it only \emph{seems} to give a
violation. The model does not violate the inequality; it is true that
the model gives the correlation
\begin{equation}
	\label{eq:cosine-expectation}
	E\big(A(\phi_A)B(\phi_B)\big|
	\text{coinc. for }A(\phi_A)\text{ and }B(\phi_B)\big)=\cos(\phi_A+\phi_B),
\end{equation}
but this is a \emph{conditional} expectation of the type used in
Theorem\nobreakspace \ref {thm:inefficiency}. We have stumbled upon a loophole that arises from the
fact that we need to postselect coincident
events, which means that we have to discard 50\% of the events right away. This
\enquote{postselection loophole} gives us a case that is similar to the
detection loophole discussed in Theorem\nobreakspace \ref {thm:inefficiency}. More accurately, we the
following theorem from~\cite{LarssonGill} can be established:
\begin{figure*}[t]
	\centering
	\includegraphics{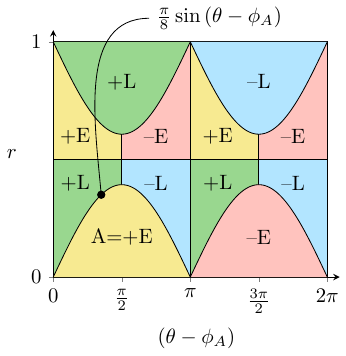}
	\includegraphics{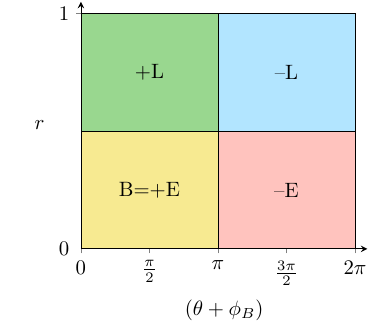}
	\caption{The local hidden variable model of \citeauthor{AKLZ}~\cite{AKLZ}. The
	hidden variable is a pair of numbers $\lambda=(\theta,r)$ evenly
	distributed over the rectangle
	$\Lambda=\{(\theta,r):0\le\theta<2\pi,0\le r<1\}$. The outcomes
	are determined by the above graphs where detections are \enquote{early}
	(E) or \enquote{late} (L). This model reproduces the quantum predictions
	for the Franson interferometric setup, including those for the
coincident detections.}
\label{fig:lhv}
\end{figure*}

\begin{theorem}\label{thm:delay}
	A local realist model with delays has the
	properties \enquote{Realism} and \enquote{Locality} from
	Theorem\nobreakspace \ref {thm:inefficiency} together with the following property:

	\begin{description}
		\item[Delays.] Detections occur after local realist time delays,
			\begin{equation*}
				T_A(\phi_A):\Lambda\mapsto\mathbb R\quad\Text{and}\quad
				T_B(\phi_B):\Lambda\mapsto\mathbb R,
			\end{equation*}
			and usage of a coincidence window gives an apparent efficiency of
			\begin{equation*}
				\eta=\underset{\parbox{14mm}{\centering\scriptsize{\upshape
				settings\\local sites}}}{\min} P\Big(\Text{coincidence}
				\Big|\Text{local detection}\Big).
			\end{equation*}
	\end{description}
	The outcomes from a local realist model with delays obey
	\begin{equation}\label{eq:bell-delay}
		\begin{split}
			\Big|E&\big(A_1B_1\big|\Text{coinc. for }A_1\Text{ and }B_1\big)
			+E\big(A_1B_2\big|\Text{coinc. for }A_1\Text{ and }B_2\big) \Big|\\
			&+\Big|E\big(A_2B_1\big|\Text{coinc. for }A_2\Text{ and }B_1\big)
			-E\big(A_2B_2\big|\Text{coinc. for }A_2\Text{ and }B_2\big) \Big|
			\le\frac6\eta-4.
		\end{split}
	\end{equation}
\end{theorem}

In the Franson setup the delays are not continuously distributed but rather
limited to two values: \enquote{Early} and \enquote{Late}. Coincidence
only occurs when these timeslots equal between the analysis stations, corresponding to a
coincidence window smaller than the delay. In inequality~(\ref {eq:bell-delay}) the bound for
quantum-mechanical violation is $\eta=3-3/\sqrt2\approx87.87\%$, which is higher
than the standard bound. The apparent efficiency that arises in the Franson
interferometer due to postselection is 50\% even with ideal detectors. It is
therefore possible for a local realist model (like the one in figure\nobreakspace \ref {fig:lhv}) to
violate inequality~(\ref {eq:bell-delay}).

At this point we want to re-establish a violation. To do this we must first
understand why Theorems\nobreakspace  \ref {thm:localrealism} to\nobreakspace  \ref {thm:delay}  fail to
provide a usable Bell test. What is the basic reason for the existence of the
model in figure\nobreakspace \ref {fig:lhv}? It is tempting to blame only the postselection, but
this is not the whole story. It should be noted that the theorems treat the
individual sites as black boxes; feed them a setting and they give an outcome in
the late or early timeslot with a value of $+1$ or $-1$, much like the setup of
figure\nobreakspace \ref {fig:epr-bohm}. We will see that taking more properties of the setup into
account will enable a violation, and the remainder of this paper will discuss
the possible ways to do this.

\section{Path Realism}\label{sec:pathrealism}

The key ingredient of the local realist model in figure\nobreakspace \ref {fig:lhv} is
that the timeslot in which detection occurs at an analysis station depends on the relation between the
hidden variables $(\theta,r)$ and the local setting ($\phi_A$ or $\phi_B$)
at the analysis station. To avoid this, one possibility is to have the \enquote{path
taken by the photon} as a realist property~\cite{Franson00,
Franson09}, in essence requiring particle-like properties of the
photons. In this case, a photon will encounter the first beamsplitter
before the variable phase modulation, so that the \enquote{decision} to \enquote{take
the long path or the short path} must be independent of the phase
modulation setting of the analysis station---the \emph{local} measurement
chosen---in contrast to a standard Bell experiment where only
independence from the \emph{remote} measurement choice is required.

It is certainly possible to list path realism as an expected model property and
test it. It is however important to note that, in this setup, path realism is
very different from measurement-outcome realism. The outcomes are EPR elements
of reality because they can be remotely predicted, and by locality they can be
argued to exist independently of what remote measurement was made. On the other
hand, the path taken is not an EPR element of reality because it cannot be
remotely predicted. There is no measurement at one site that enables a path
prediction for the remote site in the setup of figure\nobreakspace \ref {fig:franson}.

One may attempt to argue for path realism by bringing in properties from
classical physics into the picture~\cite{Franson00}, which would make models
like that in figure\nobreakspace \ref {fig:lhv} inconsistent. However, the question at hand is not
if an experiment like the above can be described within classical physics; there
is no doubt that this is not possible. Instead, the question is whether the
quantum-mechanical model can be considered complete. It is entirely possible
that our classical intuition fails us, while quantum mechanics still can be
completed. The discussion on EPR-Bell arguments is an attempt to find precisely
what minimal requirements are needed to give a contradiction with quantum
predictions. This question cannot be answered if the model is required to obey
some complicated requirements from classical physics.

Another argument for path realism~\cite{Franson09} would be to appeal
to local prediction rather than prediction from the remote site as
EPR do. By measuring in one path and finding the photon there it is possible to
predict that it is not present in the other. Locality and spacelike
separation between the paths could then be used as support for path
realism, and one could even attempt to extend the notion of
reality~\cite{Franson09} by changing \enquote{without in any way disturbing a
system} into \enquote{spacelike separation}. This is a large modification of EPR
elements of reality because EPR requires that a photon---when detected in one path---is
a different undisturbed system when predicted not to be present in the other
path, which is clearly not the case. It is true that finding the photon in one
position enables a prediction that it is not anywhere else. But this cannot be
used as evidence of an underlying realist model. Prediction of properties of
\emph{one system} immediately after measurement merely suggests that the
measurement is repeatable.

Furthermore, it is central in the EPR reasoning that the system that we predict
properties for is unaffected by the measurement used for predictions. EPR could
choose to remotely predict position or momentum and therefore conclude the
simultaneous reality of both. However, photon path is only available through a
measurement in the local analysis station, and such a measurement prevents the
remote prediction of the analysis station output, since the interference is
destroyed in the process. This means that we cannot conclude in the same way
that path and analysis station output both simultaneously are realist properties.
Even though it is claimed in Refs.~\cite{Franson00, Franson09} that
\enquote{path taken} must be a realist property independently of whether the
path measurement is performed or not, this is clearly not supported by EPR-style
reasoning.

If we choose to use path realism we must remember that path realism
on its own does not prohibit interference. First order interference is not
prohibited, since the model could randomly select a path for the photon and send
an \enquote{empty wave} through the other path, that registers any phase shift
in that path. The phase shift difference can be used to determine through which
output port to emit the photon from the second beamsplitter. Second order
interference in the present setup is also not directly prohibited by path
realism itself, because the same mechanism would work.

However, large second order interference will be prohibited by the combination
of path realism and local realist analysis station output. These two together
will enable Theorem\nobreakspace \ref {thm:localrealism} separately both for Late-Late coincidences
(both photons \enquote{taking the long path}) and Early-Early coincidences (both
photons \enquote{taking the short path}), because the existence of a delay cannot depend on
the local choice of phase modulator settings. This re-establishes the bound on the
whole set of coincidences~\cite{Franson00}.

\begin{theorem}\label{thm:path-realism}
	A local realist model with path realism has the
	properties \enquote{Realism} and \enquote{Locality} from Theorem\nobreakspace \ref {thm:localrealism}, and
	\begin{description}
		\item[Path realism.] The path taken is a realist property, and is
			setting-independent.
	\end{description}
	The outcomes from a local realist model with path realism obey
	\begin{equation}\label{eq:bell-path-realism}
		\begin{split}
			\Big|E&\big(A_1B_1\big|\Text{coinc. for }A_1\Text{ and }B_1\big)
			+E\big(A_1B_2\big|\Text{coinc. for }A_1\Text{ and }B_1\big)\Big|\\
			&+\Big|E\big(A_2B_1\big|\Text{coinc. for }A_1\Text{ and }B_1\big)
			-E\big(A_2B_2\big|\Text{coinc. for }A_1\Text{ and }B_1\big)\Big|
			\le2.
		\end{split}
	\end{equation}
\end{theorem}

Of course, path realism requires intimate knowledge of the internals of the
analysis stations, for example, that there really are two distinct paths used. This
means that path realism makes it difficult to argue for device independence
since the requirements used involve low-level properties of the devices. Also,
an experiment meant to test the inequality needs to ensure that the
\enquote{decision} to \enquote{take the long or the short path} really is
independent of the phase modulator setting. This would be accomplished by space-like
separation of the choice of phase modulation and the \enquote{photon's path choice},
that is, the event of \enquote{the photon passing the beamsplitter}. Since the
experimenters do not know the detection moments in advance, this must be done by
switching the settings fast enough to ensure that a new random setting is chosen
in-between these two events. This is in contrast to a Bell experiment where the
setting choices only need to be chosen at spacelike separation from the emission
and each other, which is a much less demanding task.

Using path realism does have benefits, since postselection has no negative
effects on the bound. The appropriate inequality has the standard bound as in
inequality~(\ref {eq:bell-path-realism}) and is violated by the quantum prediction
in equation\nobreakspace \textup {(\ref {eq:franson-qm})}. There is no local realist model with path
realism that gives the probabilities predicted by quantum mechanics for the
Franson interferometer.

\section{Local realism only}\label{sec:localrealism}

It is now interesting to ask whether it is possible to obtain a
violation using \emph{only} EPR elements of reality. Obviously, local
realist measurement outcomes is not enough. It \emph{is} possible to
establish another EPR element of reality in this setup, but it is
slightly different than the path realism used above. By replacing one of the
analysis stations with a single detector it is possible to measure the source
emission time. This enables prediction of the emission moment for
the remote photon, without in any way disturbing it. Therefore,
\emph{the moment of emission is an EPR element of reality}, which
means that a local realist model must be similar to the one in
figure\nobreakspace \ref {fig:lhv} in that it needs to specify whether a
detection is \enquote{early} or \enquote{late}. This is not the same as specifying
the path because it does not require particle-like behaviour of the
quantum object (the photon \enquote{takes a path}). The only
requirement is that the detection occurs in one of the two timeslots,
treating the analysis station like a black box
just as it is in the original Bohm-Bell setup. The important
observation here is that the detection moment also is an element of
reality and must be present in any local realist model.

In this situation, the settings $\phi_A$ and $\phi_B$ can still influence
whether a delay occurs or not. The event that needs to be spacelike
separated from the setting choices is no longer \enquote{passing the first
beamsplitter}, a point in time which here is EPR element of reality, but
the event of detection. This is because the detection is
the event that gets delayed (or not) in the local realist model. The
problem is that detection takes place after the photon has (or could
have) passed the phase modulation, so it seems impossible to have spacelike
separation between setting choice and detection event.

\begin{figure}[t]
	\centering
        \parbox[t]{2em}{\textbf{A}}%
        \parbox[t]{.4\textwidth}{\null\rule{0pt}{3cm}\includegraphics[width=\linewidth]{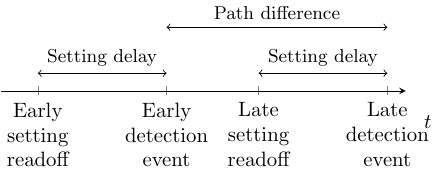}}
        \parbox[t]{2em}{\textbf{B}}
        \parbox[t]{.4\textwidth}{\null\includegraphics[width=.4\textwidth]{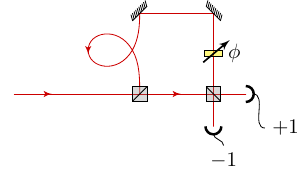}}
        \caption{\textbf{A:} Two settings influence the
          outcome, one for the early detection, and one for the late
          detection. The setting delay is the time that it takes for
          the information to travel from the phase modulator to the
          detectors, via the light path. \textbf{B:} To ensure the
          events occur in the order indicated in a), three things can
          be done: \inlinelist{ \protect\item making the path length
            difference large, \protect\item placing the phase
            modulator late in the analysis station, and \protect\item
            placing the detectors close to the second beamsplitter.}}
        \label{fig:timeline}
\end{figure}

Fortunately, there are two possible detection events, one early and
one late. This means that it is possible to have two different
settings for the two detection events: one at the \enquote{early-setting
readoff} event for the early detection, and one at the \enquote{late-setting
readoff} event for the late detection (see figure\nobreakspace \ref {fig:timeline}A). The
choice of the early setting cannot be spacelike separated from any of
the possible detection events since they are both inside (or on) the
forward light cone with respect to it. The choice of the late setting,
however, can be spacelike separated or even inside the forward light
cone of the (possible) early detection event. To make this possible,
one needs to make the path difference of the analysis station longer
than the distance from the phase modulator to the detectors. In other
words, one needs to make the path difference large, place the phase
modulator as late as possible in the analysis station and have the
detectors close to the second beamsplitter (see figure\nobreakspace \ref {fig:timeline}B).

When this is the done, the early setting can still be used by a
(hypothetical) model to delay the detection. But as soon as a delay has occured
the setting cannot cause the delay to be undone as this would violate causality. The early
detection event would already have taken place when the
late-setting choice is made. So while the early coincidences only are bounded by the
trivial bound 4 in Theorem\nobreakspace \ref {thm:localrealism} applies
for the late coincidences. We therefore have the following theorem~\cite{AKLZ}:

\begin{theorem}\label{thm:local-realism-only}
	A local realist model with long realist delays has the properties
	\enquote{Realism} and \enquote{Locality} from Theorem\nobreakspace \ref {thm:localrealism}, and

	\begin{description}
		\item[Long realist time delays.] The delay is a realist property as
			in Theorem\nobreakspace \ref {thm:delay}	and is long enough to ensure that the setting
			relevant for the late detection cannot undo the delay.
	\end{description}
	The outcomes from a local realist model with long
	realist delays obey
	\begin{equation}\label{eq:local-realism-only}
		\begin{split}
			\Big|E&\big(A_1B_1\big|\Text{coinc. for }A_1\Text{ and }B_1\big)
			+E\big(A_1B_2\big|\Text{coinc. for }A_1\Text{ and }B_2\big)\Big|\\
			&+\Big|E\big(A_2B_1\big|\Text{coinc. for }A_2\Text{ and }B_1\big)
			-E\big(A_2B_2\big|\Text{coinc. for }A_2\Text{ and }B_2\big)\Big|
			\le3.
		\end{split}
	\end{equation}
\end{theorem}

The bound in inequality~(\ref {eq:local-realism-only}) is 3. This stems from the fact that
early and late coincidences are equally probable which lets us take the mean
value of the trivial bounds 4 (early coincidences) and 2 (late coincidences).
Unfortunately, this is larger than the maximal quantum prediction $2\sqrt2$. To
establish a better bound we need to use so-called \enquote{chained} Bell
inequalities~\cite{Pearle,Braunstein90} with more terms, and this gives the following
theorem~\cite{AKLZ}.

\begin{theorem}
	The outcomes from a local realist model with long
	realist delays (as specified in Theorem\nobreakspace \ref {thm:local-realism-only}) obey
	\begin{equation}
		\label{eq:braunstein-caves}
		\begin{split}
			\Big|E&\big(A_1B_3\big|\Text{coinc. for }A_1\Text{ and }B_3\big)
			+E\big(A_1B_2\big|\Text{coinc. for }A_1\Text{ and }B_2\big)\Big|\\
			&+\Big|E\big(A_2B_2\big|\Text{coinc. for }A_2\Text{ and }B_2\big)
			+E\big(A_2B_1\big|\Text{coinc. for }A_2\Text{ and }B_1\big)\Big|\\
			&+\Big|E\big(A_3B_1\big|\Text{coinc. for }A_3\Text{ and }B_1\big)
			-E\big(A_3B_3\big|\Text{coinc. for }A_3\Text{ and }B_3\big)\Big|
			\le5.
		\end{split}
	\end{equation}
\end{theorem}

Again the bound is the mean value of the trivial bound 6 for the early
coincidences and the chained-Bell bound 4 for the late
coincidences. While the postselection does have an effect on the
bound, this case allows quantum mechanics to give a violation;
choosing the six directions $\pi/6$ apart in a plane in the order
$b_3$, $a_1$, $b_2$, $a_2$, $b_1$, $a_3$, will
yield the quantum prediction
\begin{equation}\label{eq:braunstein-caves-qm}
	\begin{split}
		\Big|\big\langle&
		A_1B_3\big|\Text{coinc. for }A_1\Text{ and }B_3\big\rangle +\big\langle
		A_1B_2\big|\Text{coinc. for }A_1\Text{ and }B_2\big\rangle \Big|\\
		&+\Big|\big\langle
		A_2B_2\big|\Text{coinc. for }A_2\Text{ and }B_2\big\rangle +\big\langle
		A_2B_1\big|\Text{coinc. for }A_2\Text{ and }B_1\big\rangle \Big|\\
		&+\Big|\big\langle
		A_3B_1\big|\Text{coinc. for }A_3\Text{ and }B_1\big\rangle -\big\langle
		A_3B_3\big|\Text{coinc. for }A_3\Text{ and }B_3\big\rangle \Big|\\
		&=6\cos\tfrac\pi6\approx 5.196,
	\end{split}
\end{equation}

There is no local realist model with setting-independent delays that gives this
value and we therefore have established a test for the Franson interferometer
that is free of the postselection loophole. Performing this experiment is however
more difficult than the one in Theorem\nobreakspace \ref {thm:localrealism} since the value in
equation\nobreakspace \textup {(\ref {eq:braunstein-caves-qm})} is so close to the bound in
inequality~(\ref {eq:braunstein-caves}). The visibility needs to be as high as 96.2\% which
is demanding. By adding even more terms to inequality~(\ref {eq:braunstein-caves}) it is
possible to slightly decrease this requirement to $94.6\%$. This minima is found
when using ten terms, increasing the number of terms even more only increases
the critical visibility. Table\nobreakspace \ref {tab:critical-visibility} lists visibility
requirements for different number of terms.

When only using local realism it becomes easier to argue for device independence
since one does not rely on internal properties of the analysis stations. It is,
of course, still important to verify that the delays do occur at equal
probability, but it is no longer needed to verify the existence of two distinct
paths. Similarly as before, an experiment meant to test the inequality needs to
ensure that the possible early detection event really is independent of the late
phase delay choice. And again, since the experimenters do not know the detection
moments in advance, this must be done by switching the settings fast enough to
ensure that a new random setting is chosen in-between the early detection event
and the late-setting readoff. This is similarly difficult to the corresponding
requirement when using path realism (see above), but this is the price to pay
for only using EPR elements of reality in the derivation of the bound. One
should also note that the early-setting choice and the late-setting choice needs
to be independent, which would be achieved with a good source of randomness. For
the purists~\cite{Scheidl-etal10} the late-setting and early-setting choices
need to be spacelike separated, something that can be achieved with independent
sources of randomness suitably arranged around the analysis stations.

\begin{table}
	\centering
	\begin{tabular}{cccc}
		Number&Emission-time&Quantum&Critical\\
		of terms&realism bound&prediction&visibility\\
		\midrule
		4&3&$4\cos\frac\pi4\approx2.828$&$>100\%$\\
		6&5&$6\cos\frac\pi6\approx5.196$&$96.23\%$\\
		8&7&$8\cos\frac\pi8\approx7.391$&$94.71\%$\\
		10&9&$10\cos\frac\pi{10}\approx9.511$&$94.63\%$\\
		12&11&$12\cos\frac\pi{12}\approx11.59$&$94.90\%$\\
		$2N\ge14$&$2N-1$&$2N\cos\frac\pi {2N}$&incr.\ with $N$\\
	\end{tabular}
	\centering
	\caption{Critical visibilities for violation of chained Bell
		inequalities that the outcomes from a local realist model with long
		realist delays (as specified in Theorem\nobreakspace \ref {thm:delay}) must
	obey.}\label{tab:critical-visibility}
\end{table}

We have found that when establishing the time of emission as an EPR
element of reality, postselection still has negative effects, but only
on half of the selected subensemble. This modifies the relevant Bell
inequalities, but some are still violated by the quantum prediction in
equation\nobreakspace \textup {(\ref {eq:cosine-correlation})}. There is no local realist model with realist emission
time that gives the probabilities predicted by quantum mechanics for
the Franson interferometer.

\section{Modified setups}\label{sec:modified-setups}

Another approach to reach a violation of local realism from the
quantum-mechanical predictions is to modify the setup. A number of
alternatives exist, and we will here briefly go through three of these
alternatives.

\begin{figure*}[p]
	\centering
	\includegraphics[width=\linewidth]{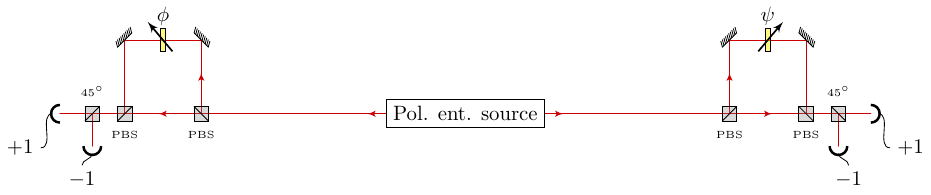}
	\caption{Using a polarization-entangled source. The time-correlated
		photons are still sent out at unknown moments in time, but are now
		also polarization-entangled. The beamsplitters used are polarizing
		beamsplitters (PBS), and the interference occurs at the third PBS
		at each site, because this is oriented $\pi/4$ in relation to the
		other two. In this setup there is no postselection, so that
		Theorem\nobreakspace \ref {thm:localrealism} can be used directly, and the bound is violated by the
	quantum prediction.}\label{fig:pol-ent}
\end{figure*}
\begin{figure*}[p]
	\centering
	\includegraphics[width=\linewidth]{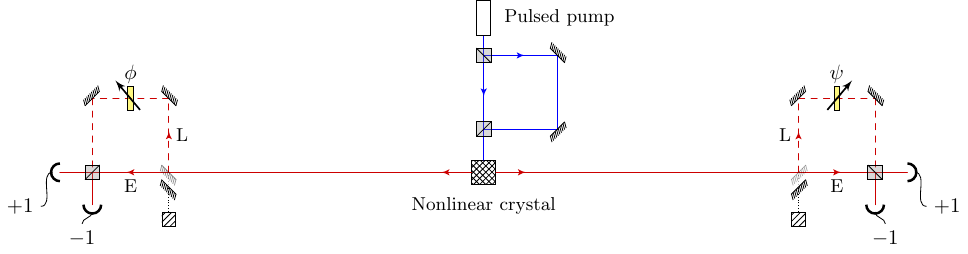}
	\caption{\label{fig:controlled-mirrors}	Time-bin entanglement. The
	mirrors are synchronized with the source so that photons produced by the
	early part of the pulse would be delayed in the analysis station, while
	photons produced by the late part of the pulse would not be delayed. In
	this setup there is no postselection, so that Theorem\nobreakspace \ref {thm:localrealism} can be
	used directly, and the bound is violated by the quantum prediction. If we
	instead replace the controlled mirrors with passive beamsplitters we have
	phase-time encoding which gives us Theorem\nobreakspace \ref {thm:local-realism-only}.}
\end{figure*}
\begin{figure*}[p]
	\centering
	\includegraphics[width=\linewidth]{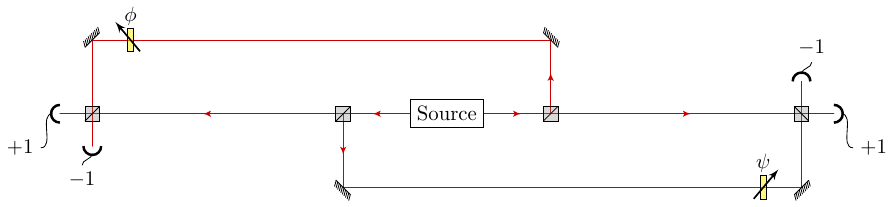}
	\caption{Hugging interferometers. This setup has many of the
		properties of the original Franson setup. The interferometers are
		very large and this becomes experimentally challenging.
		Here, the path taken by each photon is an EPR element of reality
		so that Theorem\nobreakspace \ref {thm:localrealism} can be used, and the bound is violated by the
	quantum prediction.}\label{fig:genuine}
\end{figure*}

The first modified setup was proposed by \citeauthor{Strekalov96} in 1996~\cite{Strekalov96}
(see~figure\nobreakspace \ref {fig:pol-ent}). This setup uses a polarization-entangled source
and three polarizing beamsplitters at each site. The interference occurs at
the third and last polarizing beamsplitter. In this setup, the path taken
\emph{is} an EPR element of reality because a polarization measurement (in
place of one analysis station) can be used to remotely predict which path the
remote photon is going to travel through. But this is not needed; because of
the polarization entanglement the photon pairs always end up in the same
timeslot, giving coincident detection. There is no
postselection, and we can in fact use Theorem\nobreakspace \ref {thm:localrealism} (modulo other
experimental problems). This is good, because the experimental realization
proposed in~\cite{Strekalov96} does not use separate paths, but instead uses
a single birefringent optical element at each site to implement the whole
analysis station, so an argument based on a realist path cannot be
used. But as we have seen, local realism can be violated in this setup even
when the paths coincide.

The second alternative, sometimes called \enquote{time-bin entanglement}, uses a
source proposed in \citeauthor{Brendel99} in 1999~\cite{Brendel99} and
switched mirrors in place of the first beamsplitters~\cite{Tittel99}. The
source uses a pulsed pump, an unbalanced interferometer, and a nonlinear
device that creates photon pairs. The active mirrors are pushed in and
pulled out of the photon path in sync with the source (see
figure\nobreakspace \ref {fig:controlled-mirrors}). In this setup, the path taken is also an EPR
element of reality, because a measurement of the time of emission (in place
of one analysis station) can be used to remotely predict which path the remote
photon is going to travel through. Again, this is not needed; because of the
synchronization of the mirror positions the photon pair is detected in the same
timeslot and every pair gives coincident detections. There is no
postselection, and Theorem\nobreakspace \ref {thm:localrealism} (modulo other experimental
problems) can be used to rule out local realist models.

It is possible to modify the time-bin-entanglement setup further by using
passive switching. \Citeauthor{Gisin2002} refers to this setup as
\enquote{phase-time encoding}~\cite{Gisin2002}. In this method, the movable
mirrors are replaced by beam splitters. We can remove one of the analysis
stations and measure the emission time from the source, which becomes an EPR
element of reality. However, in contrast to time-bin-entanglement it is not
possible to predict the path taken and we get a situation very similar to the
Franson interferometer. In fact, with passive switching we recover the weaker
Theorem\nobreakspace \ref {thm:local-realism-only} and the only difference to the Franson setup is
the fact that phase-time encoding uses a pumped source.

The third alternative in this list was proposed by \citeauthor{Cabello09} in
2009~\cite{Cabello09} and uses interferometers in a \enquote{hugging}
configuration as seen in figure\nobreakspace \ref {fig:genuine}. In this setup, postselection
is performed because the two photons both may end up at the same site, at
both input ports of the beamsplitter at that site but in different timeslots.
The events that give coincident detection at both sites make
up 50\% of the total. But also here the path is an EPR element of reality,
because a local path measurement (done by removing the final beamsplitter)
can be used to remotely predict which path the remote photon is going to
emerge from. Since the path is an EPR element of reality, we can use
Theorem\nobreakspace \ref {thm:path-realism} (modulo other experimental problems), and the
quantum prediction will violate the bound. There is no local realist model
(with realist path) that gives the quantum predictions of this setup, and
because of this, the hugging interferometer setup is sometimes called a
setup with \enquote{genuine energy-time entanglement}.
\Citeauthor{Lima2010}~\cite{Lima2010} and \citeauthor{Cuevas2013}~\cite{Cuevas2013}
have both demonstrated violations of Bell's inequality using hugging
interferometers, the latter was performed with a fibre length of
\SI{1}{\kilo\meter}.

\section{Conclusions}\label{sec:conclusions}

This paper has discussed tests of local realism using energy-time
entanglement. Even though most of the proposed experiments use
postselection, we have seen that certain types of models can be ruled
out. But we have also seen that the tests are subtle, because
depending on the precise requirements made on the local realist model,
the same experimental data can imply
\inlinelist{
\item no violation,
\item smaller violation than usual, or
\item full violation of the appropriate	statistical bound.
}

Using only that the measurement outcomes are EPR elements of reality
is not enough to yield a violation~\cite{AKLZ}: the appropriate Bell
inequality would be trivial and reads
\begin{equation}\label{eq:trivial-inequality}
	\begin{split}
		\Big|E&\big(A_1B_1\big|\Text{coinc. for }A_1\Text{ and }B_1\big)
		+E\big(A_1B_2\big|\Text{coinc. for }A_1\Text{ and }B_2\big)\Big|\\
		&+\Big|E\big(A_2B_1\big|\Text{coinc. for }A_2\Text{ and }B_1\big)
		-E\big(A_2B_2\big|\Text{coinc. for }A_2\Text{ and }B_2\big)\Big|
		\le4.
	\end{split}
\end{equation}

Adding that the moment of emission is an EPR element of reality
enables a violation, although a weaker violation than that of the
usual Bell setup~\cite{AKLZ}: there is a violation using the chained
equation\nobreakspace \textup {(\ref {eq:braunstein-caves-qm})} with six terms while the appropriate
four-term Bell inequality would read
\begin{equation}\label{eq:element-of-reality}
	\begin{split}
		\Big|E&\big(A_1B_1\big|\Text{coinc. for }A_1\Text{ and }B_1\big)
		+E\big(A_1B_2\big|\Text{coinc. for }A_1\Text{ and }B_2\big)\Big|\\
		&+\Big|E\big(A_2B_1\big|\Text{coinc. for }A_2\Text{ and }B_1\big)
		-E\big(A_2B_2\big|\Text{coinc. for }A_2\Text{ and }B_2\big)\Big|
		\le3.
	\end{split}
\end{equation}

Requiring that the model also uses a realist path will enable a
violation, equally large as that from a standard Bell
test~\cite{Franson00,Franson09}. Note that, in the original Franson setup,
the path is not an EPR element of reality; it cannot be predicted
without disturbing the system. But path realism can still be listed
as an expected model property, and tested in experiment; one motive
for requiring this particle-like behaviour from the model would be
correspondence to properties from classical physics. The violation
would be equally strong as that of the usual Bell setup: the
appropriate Bell inequality has the usual bound
\begin{equation}\label{eq:original-bell}
	\begin{split}
		\Big|E&\big(A_1B_1\big|\Text{coinc. for }A_1\Text{ and }B_1\big)
		+E\big(A_1B_2\big|\Text{coinc. for }A_1\Text{ and }B_2\big)\Big|\\
		&+\Big|E\big(A_2B_1\big|\Text{coinc. for }A_2\Text{ and }B_1\big)
		-E\big(A_2B_2\big|\Text{coinc. for }A_2\Text{ and }B_2\big)\Big|
		\le2.
	\end{split}
\end{equation}

A final alternative is to modify the setup to ensure that the path
taken is an EPR element of reality~\cite{Strekalov96, Brendel99,
Tittel99,Cabello09}. This enables a violation that is equally strong
as that of the usual Bell setup: the appropriate Bell inequality is
inequality~(\ref {eq:original-bell}).

These considerations must also be taken into account in quantum
communication. For example, in Bell-inequality based quantum
cryptography~\cite{Ekert91}, the inequality is used as test of
security. The original Bell inequality (bounded by 2) is only
available as a security test if path realism can be used, and this is
only possible when
\inlinelist{
\item there really are distinct paths within the
	analysis stations, and
\item the attacker is unable to control which path
	the \enquote{photon will take}.
}
This is highly device-dependent; when
aiming for device-independent security, the security test should not
rely on the internal structure of the analyzing stations. Thus, path
realism should not be used. The users must rely on the black-box
formulation obtained when using emission time as an EPR element of
reality. In this case, the original four-correlation Bell inequality
cannot be used as test of security (see above), and using the chained
inequalities, the security margin will be smaller than from the
standard Bell test, since the critical visibility is higher than in
this test.

Energy-time entanglement obtained with the Franson interferometer and
its variants is a subtle way to test local realistic models, with or
without added properties. It will be used in many future experiments
intended to violate local realism, but in performing them it is
important to be aware of exactly what is tested in these experiments,
and the size of the violation. Even with the subtleties associated
with this interferometer, or more correctly, because of these
subtleties, the interferometer will continue to be an important tool
to extend our knowledge on the foundations of quantum mechanics.

\ack{}
The authors would like to thank J.~D.~Franson, A.~Cabello, and
M.~\.Zukowski for interesting and stimulating discussions.

\printbibliography{}

\end{document}